# Recent advances in Ti and Nb explosion welding with stainless steel for 2K operating (ILC Program)

(presented by B.Sabirov)

JINR, Dubna, 141980, Russia: ................B.Sabirov, J.Budagov, A.Sissakian, G.Shirkov, Yu.Taran,
.                                                            G.Trubnikov
FNAL, Batavia, IL, 60510, USA:  .............N.Dhanarai, M.Foley, E.Harms, A.Klebaner, D.Mitchell,
.                                                            S.Nagaitsev, W.Soyars
RFNC, Sarov, 607188, Russia .................. V.Rybakov, Yu.Samarokov, V.Zhigalov
INFN, Pisa, 56127, Italy..............................A.Basti, F.Bedeschi

## Abstract

The world first samples 0f Ti+SS and Nb+SS joints were manufactured by an explosion welding technology demonstrating a high mechanic properties and leak absence at $4.6 \cdot 10^{-9}$ atm·cc/sec. Residual stresses in bimetallic joints resulting from explosion welding measured by neutron diffraction method are quite high (≈1000 MPa). Thermal tempering of explosion welded Ti+SS and Nb+SS specimens leads to complete relaxation of internal stresses in Ti, Nb and Stainless steel and makes the transition elements quite serviceable.

JINR group joined the ILC topic in 2006 with accepted our proposal to use the explosion welding technique for making bimetallic titanium–stainless steel (Ti + SS) and niobium–stainless steel (Nb + SS) joints for upgrade of ILC cryomodule. At the first stage the task was to make a bimetallic transition from the helium supply tube of stainless steel (SS) to the cryomodule shell of titanium [1]. This would appreciably lower the cost of the accelerator. It was a nontrivial problem because Ti and stainless steel cannot be welded by conventional welding techniques. We came into contact with experts of the Russian Federal Nuclear Centre (RFNC, Sarov), where they have mastered a method of welding dissimilar materials by explosion. The explosion welding technique has long been known, but it is mainly used to weld flat pieces [2,3]. After numerous mutual consultations the welding technology for cylindrical pieces was developed. The pilot specimen was severely tested for strength and leakage [4]. The tests were carried out in Sarov, Dubna, Pisa (Italy), and Fermilab under various shock conditions: thermal cycling in liquid nitrogen, cooling to 1.6 K, pressure up to 6 atm inside the specimen [5]. The mechanical test made using special fixture (Fig.1). Measurement of the welded joint shear strength was carried out. An impressive result is obtained: $\zeta_{sh} \approx 250$ MPa. Metal strengthening is observed in the welded joint area. The highest material strengthening occurs in a narrow area ~0.5 mm wide near the titanium-steel interface; beyond this area strengthening decreases (Fig.2). A leak check tests shown rather good result: absence of leakage at the leak detector background indication ≈$10^{-10}$ atm·cm$^3$·s$^{-1}$.

To verify the results, we manufactured and tested another 24 Ti + SS transition elements. The tests for leakage and strength of the joints yielded similar results [6].

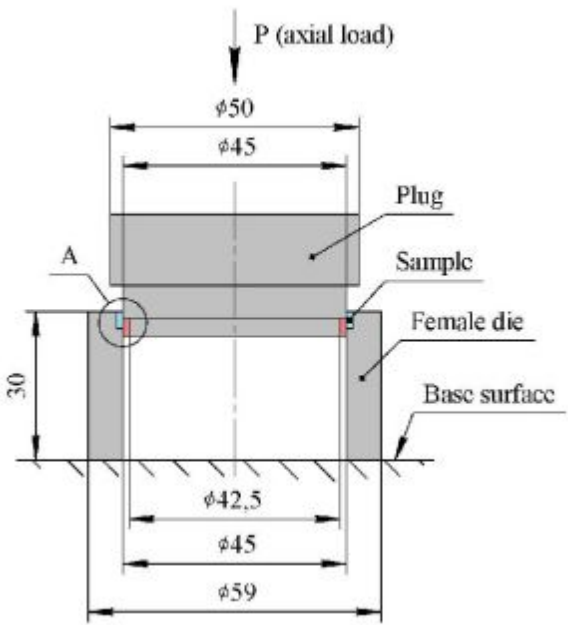 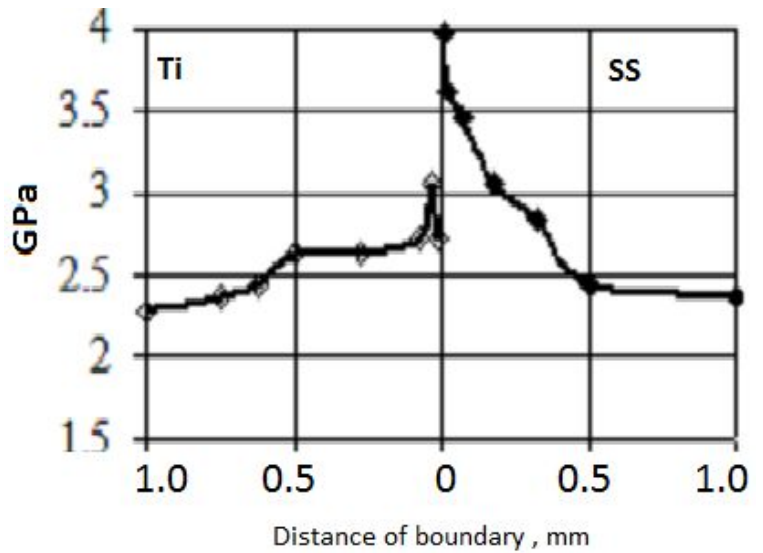

Fig.1 *Setup for shear measurement*  Fig.2 *Result of strengthening measurement of Ti and SS joint field*

Based on the experience gained with the Ti + SS specimens, we got down to solving the next problem set to us by the ILC management: to make the ILC project even cheaper, it was proposed to consider a possibility of replacing titanium with stainless steel in the shell of the helium cryostat. Thus, it was necessary to develop a transition from the stainless-steel shell to the niobium cavity—the main acceleration element of the accelerator [7]. In Sarov four Nb + SS elements were produced using two explosion

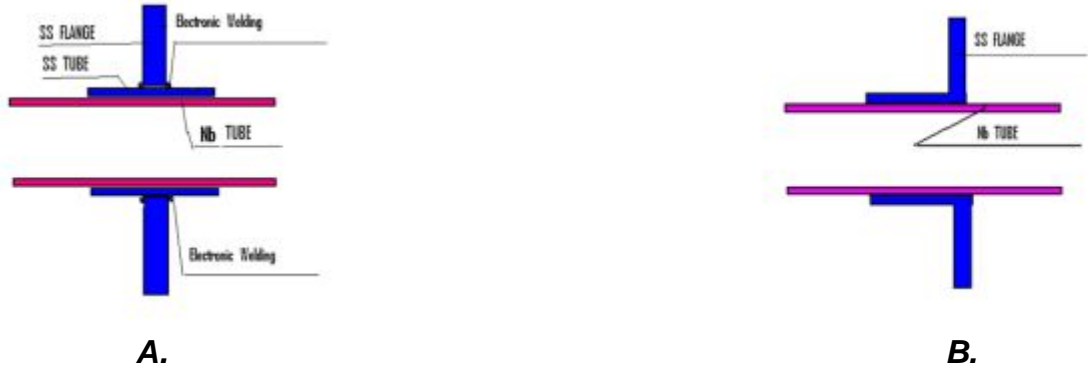

welding schemes: external (**A.**) and internal (**B.**) cladding.

In the external cladding scheme the steel tube is welded by the outside explosion to the niobium one and then the steel flange is welded to it, usually by the electron-beam welding. In the internal cladding scheme the niobium tube is welded by the inside explosion to the steel flange.

Apart from the initial problem to the make a Nb → SS transition by the explosion welding technique, we were to solve an important problem of finding out whether NB + SS transition elements made by the explosion welding could be used in generation IV cryomodules of the ILC. The point was that in our specimen the transition from the niobium cavity to the steel flange was obtained by the electron-beam welding of the niobium tube to the niobium cavity. Since the niobium melting temperature is 2460°C, the question arises as to how the Nb + SS joint will withstand this high thermal load. Preliminary leak tests by thermal cycling in liquid nitrogen carried out with the Nb–stainless steel joint in specimen #5 made by the internal cladding showed rather good results: after six thermal cycles and ultrasound cleaning no leakage was found in the specimen at the background leak detector indication $2 \cdot 10^{-9}$ atm·cm$^3$/s (gaseous He was used for vacuum testing). The decisive test should have been the test for resistance of the welded joint to high temperature. Niobium rings of the size corresponding to the size of the niobium tube in the transition element were prepared for that test. The rings were welded to the tube on both sides by electron-beam welding at the Sciaky Company (Chicago). The very first leak measurement at room temperature revealed large leaks at two points on the joint of niobium and stainless steel. The holes were stopped with the Apiezon-Q vacuum paste, and the test was continued in the thermal cycling regime. After the third cycle one more leak appeared. The test was terminated. Figure 1 shows the specimen with the leaks indicated by black arrows.

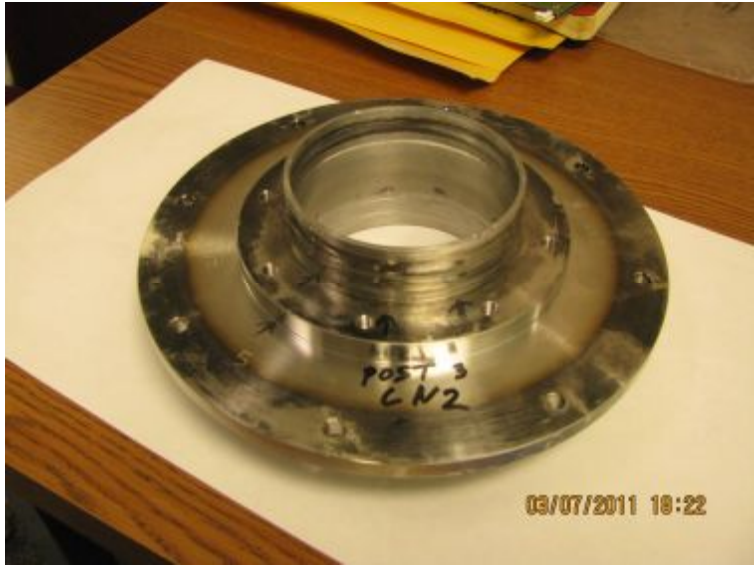

Thus, the tightness of the explosion-welded joint was shown to be greatly affected by high temperature. This appeared to arise from structural changes of materials caused by explosion welding.

Fig. 3. *Nb + SS transition element with the revealed leaks.*

It is well known that many production processes like machining, forging, stamping, rolling, welding, etc. may give rise to a strong field of residual internal stresses inside the material or product due to its plastic deformation. Residual stresses are often inhomogeneous and usually unpredictable. Destruction of mechanical components or structures results not only from stresses arising in the course of their use but also from the superposition of the former stresses on the residual stresses. Therefore, we had to solve two problems to overcome the difficulty: (i) to find out how high those internal residual stresses are and (ii) to find the way to relax those stresses.

Residual stress is traditionally measured by mechanical, physicotechnical, and diffraction methods. They are divided into three categories, destructive, semidestructive, and nondestructive. Destructive and semidestructive methods involve only a few nonrepeatable measurements, and thus the residual stress calculation may be doubtful as applied to all samples. Because of their strong absorption in most materials, X-rays and electrons are used only in surface measurements. Neutrons have an appreciably advantage as probes of the atomic structure. Unlike X-rays and electrons which interact with electron shells, neutrons interact with atomic nuclei, which may result in their diffraction. Neutrons are weakly absorbed by materials, which allows nondestructive measurements of deformation deep in samples. Several international and national scientific centres have built neutron diffraction facilities for a wide range of research.

We measured residual deformation in bimetallic Ti + SS tubes made by the explosion welding. The Ti + SS tubes were chosen because the welding technique is identical for Ti + SS and Nb + SS joints and thus the physics of diffusion of materials during the explosion and the generation of residual stresses at the joints should also be identical. Measurements were carried out with the POLDI stress diffractometer on the neutron beam from the ISIS reactor of the Paul Scherrer Institute (Switzerland) [8]. The residual stresses in the bimetallic Ti + SS tube measured during the scanning of the titanium–stainless steel joint are shown in Fig. 4.

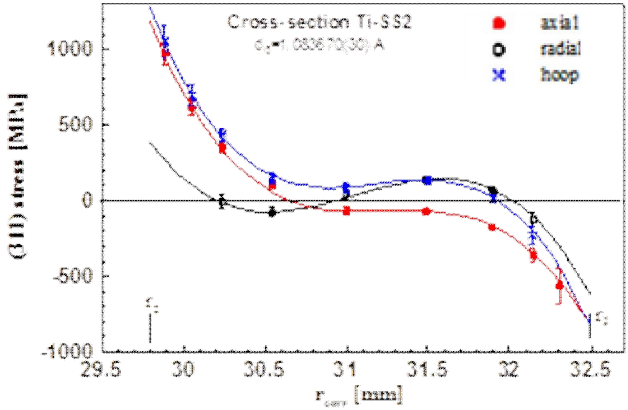

Fig. 4. *Measured (points) and fitted (curves) radial dependence of the stress tensor components obtained for the peak (311) in the Ti–SS cross section.*

As is evident from the plot, the residual stress is quite considerable, amounting to ~1000 MPa. Since additional thermal stresses may arise from electron-beam welding or deep cooling in liquid helium, their superposition can make titanium turn into the state corresponding to the deep plastic region. This may cause local microcracks in the Ti + SS (or Nb + SS) joint, which in turn may adversely affect tightness of the transition element when it is used in the cryomodule.

It was decided to subject the Nb + SS transition elements made by external cladding to thermal tempering for relaxation of residual stresses both before and after the electron-beam welding. The thermal tempering regime was agreed upon with Fermilab's leading materials scientist Lance Cooley: heating in vacuum to 750$^{o}$C at a rate of 3$^{o}$C/min, holding at this temperature for 120 minutes, and natural cooling in vacuum for a night with the heat-treating furnace off. When cooled, the specimens were subjected to ultrasound cleaning in a 2% solution of the Micro90 reagent in deionized water at room temperature for 30 minutes followed by washing in isopropyl alcohol and drying in the clean room for a night.

The leak test was carried in the thermal cycling regime using the Dupont (Ametek upgrade) leak detector with the sensitivity $10^{-10}$ atm·cc/s. The detector was calibrated using a specimen with the standard leak $5.3 \cdot 10^{-8}$ atm·cc/s. Six thermal cycles were carried out in total. The results of the measurements showed a minimum difference from cycle to cycle. The table presents the measurement results for one of the cycles.

| Cycle 2 | Specimen 1 | Specimen 2 |
|---|---|---|
| Background leak at room temperature | $1.0 \cdot 10^{-9}$ atm·cc/s | $5.3 \cdot 10^{-9}$ atm·cc/s |
| Background leak at 77 K | No change | No change |
| Specimen in polyethylene package, gaseous He injected, 77 K | $1.2 \cdot 10^{-9}$ atm·cc/s | $5.0 \cdot 10^{-9}$ atm·cc/s |
| Specimen heated to room temperature, gaseous He injected | No change | $4.9 \cdot 10^{-9}$ atm·cc/s |

The next step was the test at the temperature 2 K. The specimen was fixed to a special rig for inserting into the Vertical Test Dewar (VTD) with liquid helium.
In the figure one can see the cables from the temperature sensors attached to the Nb + SS specimen at its top, bottom, and sides.
The cooling procedure lasts for 15 to 20 hours: cooling in liquid nitrogen, supply of two-phase helium and cooling to 4 K, and finally evacuation and a change of helium to the superfluid state at the temperature 1.6 to 2 K.

One-hour long measurements with the Residual Gas Analyzer (RGS) after evacuation down to $10^{-9}$ Torr at the temperature 2 K and background level $4.6 \cdot 10^{-9}$ atm·cc/s did not reveal any leak in specimen 2. A similar result was obtained for the other specimen.

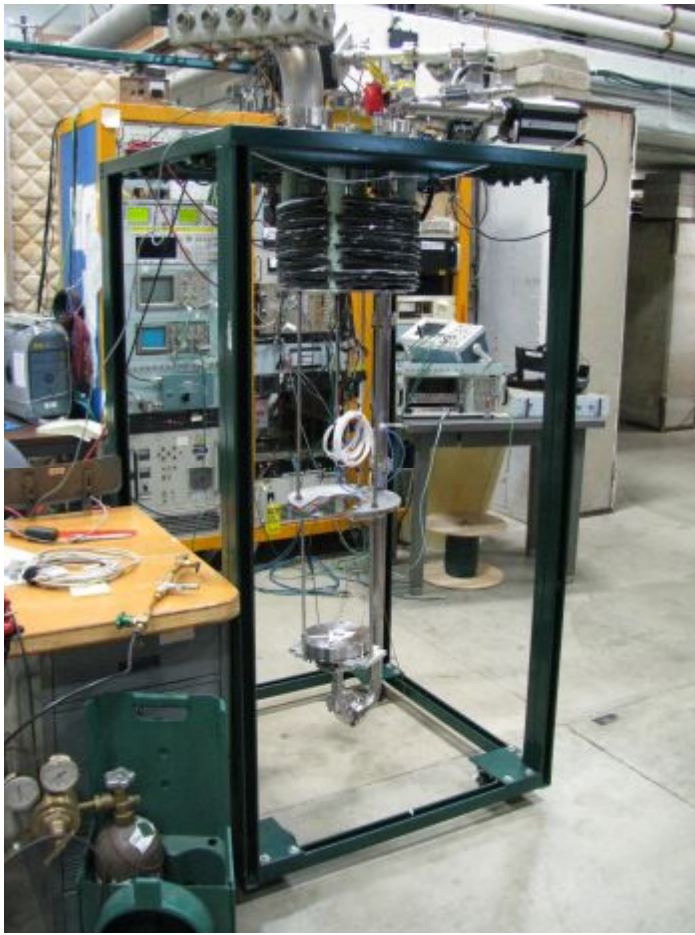

Fig. 5. *Specimen 2 assembly for insertion into the VTD.*

## Conclusion

Thus, our investigations have shown that explosion welding allows quite adequate bimetallic components to be made for cryogenic units of linear accelerators. This technique is applicable to any other cryogenic systems.
It is also shown that residual stresses resulting from explosion welding are quite high, which may cause plastic deformation and destruction of bonding of the materials. Thermal tempering of explosion-welded Ti + SS and Nb + SS specimens leads to complete relaxation of internal stresses in Ti, Nb + stainless steel joints and makes the transition elements quite serviceable.

B. Sabirov, JINR, Dubna, RUSSIA.